\documentstyle[11pt,epsfig]{article}
\newcommand{\beq}{\begin{equation}}
\newcommand{\eeq}{\end{equation}}
\newcommand{\bea}{\begin{eqnarray}}
\newcommand{\eea}{\end{eqnarray}}
\def\lsim{\raise0.3ex\hbox{$\;<$\kern-0.75em\raise-1.1ex\hbox{$\sim\;$}}}
\def\gsim{\raise0.3ex\hbox{$\;>$\kern-0.75em\raise-1.1ex\hbox{$\sim\;$}}}
 
\begin{document}
\begin{center}
{ Very High Energy Antineutrinos from Photo-disintegration of Cosmic Ray Nuclei} 

\medskip{Nayantara Gupta

Astronomy $\&$ Astrophysics, Raman Research Institute, 
Bangalore, 560080, India}
\footnote{nayan@rri.res.in}
\end{center}
\begin{abstract}
The photo-disintegration of cosmic ray nuclei by starlight leads to the production of secondary antineutrinos. We have assumed that the flux of the ultrahigh energy cosmic ray nuclei near the Galactic plane region is the same as that observed near the earth and calculated the antineutrino flux produced from their photo-disintegration.
The IceCube detector has measured the neutrino/antineutrino flux in the TeV-PeV energy range. Our calculated secondary antineutrino flux in the energy range of 10-100 TeV is found to be much less compared to the flux detected by the IceCube collaboration. The upper limit on the intensity of the radiation field in the extragalactic medium is much lower than that near the Galactic center.
If we extend our formalism to the extragalactic medium the contribution from the photo-disintegration of ultrahigh energy cosmic ray heavy nuclei remains insignificant due to their very low flux. 
\end{abstract}

Keywords: high energy cosmic rays, photo disintegration

\section{Introduction}
The cosmic ray detectors of CASA-MIA \cite{casa}, Tunka \cite{tunka}, GAMMA \cite{gamma} and KASCADE-Grande \cite{kas} have measured the cosmic ray flux from the knee region at around $10^{17}$ eV which mostly contains iron nuclei. This is known as the ``iron knee". Beyond the knee there is a change in the composition of the cosmic rays from heavy to light nuclei.

\par
The radiation field produced from the stars is most intense near the Galactic center region where the density of the stars is the highest and gradually decreases with the distance \cite{mos} from this region. The cosmic rays emitted from the Galactic and extragalactic sources interact with the interstellar radiation and matter during their propagation through the interstellar medium. 
Secondary neutrinos and gamma rays are produced from their interactions 
\cite{gupta1,gupta2,joshi,ahlers}. The pure hadronic interactions have negligible contribution to the observed neutrino events in the IceCube detector 
\cite{aart1,aart2,aart3,aart4}.

\par

The photo-disintegration of cosmic ray heavy nuclei has been studied earlier analytically \cite{stecker,hooper1,anch1} and with Monte Carlo simulations 
\cite{hooper2}. 
 More recently it has been suggested that the antineutrinos produced in the photo-disintegration \cite{anch2} of some very high injected flux of ultrahigh energy cosmic rays (UHECRs) may explain the flux observed by the IceCube neutrino detector. In photo-disintegration protons and neutrons are produced almost equally. The neutrons decay to protons and antineutrinos. 
In this paper it has been assumed that 1 to 10$\%$ of the high energy protons produced in the decay of the high energy neutrons reach us due to magnetic shielding and their flux is the same as the observed cosmic ray proton flux in the energy range of $10^{8.5}$ to $10^{9.5}$ GeV.
\par

Parametrizations of the all particle cosmic ray spectrum are given in \cite{gaisser1} for the following three population models (i) Hillas model (ii) global fit model.
 We have used their parametrizations of the diffuse cosmic ray flux to calculate the antineutrino flux produced in the photo-disintegration of cosmic ray heavy nuclei.  According to the Hillas model the Galactic cosmic ray spectrum ends at the knee which mostly originates from supernova remnants. The contribution to the diffuse cosmic ray flux from the extragalactic sources is significant at the ankle. This scenario is based on the amplification of the magnetic fields in non-linear diffusive shock accelerations which determines the maximum energy of the cosmic rays produced in Galactic SNRs. In this model at least three populations of particles are needed to explain the observed cosmic ray flux. Many more populations may be introduced to explain the cosmic ray spectrum in much more detail. In the global fit model the fluxes of the individual cosmic ray nuclei measured by the CREAM experiment \cite{ahn} are well explained. This model is also consistent with the ``iron knee'' observed by the KASCADE-Grande at $10^{17}$ eV 
\cite{antoni}. In each population all the particles have the same maximum rigidity. Another important aspect is the higher energy populations can significantly contribute to the cosmic ray flux at lower energy. For more details on these models the readers may see \cite{gaisser2}. 
\par
Although, the radiation field depends on the Galactocentric radius R and the altitude above the Galactic plane z, a uniform background radiation field is assumed in this work near the Galactic plane region.
 The strength and spectral distribution of this radiation field are assumed as given for R=0,4kpc and z=0 in Fig.1.(bottom panel) of \cite{mos}.
We have denoted the two cases by IR1 and IR2. In the first case (IR1) it is assumed that the Galactic plane region has a uniform radiation field, the same as the one given for R=0,z=0 in their paper. IR2 is the case corresponding to the radiation field equal to that given for R=4kpc,z=0. 

\section{The Cosmic Ray Flux Near the Galactic Plane:}
The cosmic ray flux and its composition near the Galactic plane region are unknown. Even near the earth the composition of the UHECRs is not yet known. The measured flux near the earth has been fitted with the Hillas model and the global fit model  earlier \cite{gaisser1}. 
The all particle spectrum (GeV$^{-1}$ m$^{-2}$ sec$^{-1}$ sr$^{-1}$) in the three population models is
\begin{equation}
\frac{dN_{A_i}(E_{A_i})}{d E_{A_i}\, dS\, dt\,d\Omega}=\Sigma_{j=1}^{3}a_{ij}E_{A_i}^{-\gamma_{ij}-1}\times exp\Big[-\frac{E_{A_i}}{Z_i R_{cj}}\Big]
\label{cosmicray_spectrum}
\end{equation}

We have used the above cosmic ray flux in this work. 
The subscript $i=1,5$ runs over the standard five groups of particles p, He, CNO, Mg-Si and Fe. The three populations are denotd by $j=1,3$. 
 Area has been denoted by $S$ and charge of a nucleus by $Z_i$. For CNO and Mg-Si we have used the mean values of their charges.
The values of the parameters are given in Table 1 and Table 2 for the Hillas model and the global fit model respectively. Note that in eqn.(\ref{cosmicray_spectrum}) the energy $E_{A_i}$ is the energy of the nucleus for the $i^{th}$ group, $E_{A_i}=A_i\,E_n$ where $A_i$ is the average mass number for the $i^{th}$ group and $E_n$ is the energy of each nucleon. 
 In the Hillas model the fluxes of the heavy nuclei are higher compared to the global fit model. The population 3 in the global fit model contains only protons and iron nuclei whereas in the Hillas model proton, He, CNO, Mg-Si and iron fluxes are almost equal. In the next section we do not use the subscript $i$ anymore to write the cosmic ray flux for the individual groups. 
\section{Secondary Antineutrinos from the Galactic Plane Region:}
 The cross-section of photo-disintegration \cite{kara} for a medium or heavy nucleus of mass number $A$ in the rest frame of the nucleus 
is
\begin{equation}
\sigma_A(\epsilon^*)=\sigma_{0A}\frac{(\epsilon^* \triangle)^2}{({\epsilon^*}^2-{\epsilon'_0}^2)^2+(\epsilon^*\triangle)^2}
\end{equation}
where the photon energy in the nuclear rest frame $\epsilon^*$ is below 30 MeV. 
Above 30 MeV the cross-section is energy independent with a smaller value of $A/8$ mb. The values of the constants are as follows, cross-section $\sigma_{0A}=1.45\,A$ mb, the central value of GDR (Giant Dipole Resonance) $\epsilon'_0=42.65\, A^{-0.21}$ MeV for mass number of nuclei $A>4$ and width of the GDR $\triangle=8$ MeV. 
The rate of photo-disintegration ($R_{ph}$) is calculated for the background radiation density per unit energy (IR) $\frac{dn(x)}{dx}$.
The Lorentz factor of each nucleon of mass $m_n$ and energy $E_n$ is $\gamma_n=E_{A}/(A \, m_n)$. 
\begin{equation}
R_{ph}=\frac{c}{2{\gamma_n}^2}\int_{\epsilon_{th}}^{2\gamma_n \epsilon_{max}}
 \epsilon^* \sigma_A(\epsilon^*) d\epsilon^*\int_{\epsilon^*/2\gamma_n}^{\epsilon_{max}} \frac{dn(x)}{dx}\frac{dx}{x^2}.
\label{rate_photd}
\end{equation}
The threshold energy of photo-disintegration in the rest frame of the nucleus is $\epsilon_{th}\sim 2$ MeV and the maximum energy of the photons in the background radiation field is $\epsilon_{max}$.

Eq.(\ref{rate_photd}) is simplified after including the expression for the total cross-section of photo-disintegration. The cross-section corresponding to the GDR can be approximated by a delta function \cite{anch1}.
\begin{equation}
\sigma_A(\epsilon^*)=\pi \sigma_{0A}\frac{\triangle}{2}\delta(\epsilon^*-\epsilon'_0)
\end{equation}
The expression for the rate of photo-disintegration in the case of GDR is simplified to
\begin{equation}
R_{ph,GDR}=\frac{c\,\pi\, \sigma_{0A}\, \epsilon'_0\, \triangle}{4\, \gamma_n^2}
\int_{\epsilon'_0/2\gamma_n}^{\infty} \frac{dn(x)}{dx} \frac{dx}{x^2}.
\label{dis_rate}
\end{equation}
 
 The constant cross-section of photo-disintegration above 30 MeV gives
\begin{equation}
R_{ph,const}=\frac{Ac}{16\gamma_n^2}\int_{30 MeV}^{2\gamma_n\epsilon_{max}}
\epsilon^* d\epsilon^*\int_{(\epsilon^*/2\gamma_n)}^{\epsilon_{max}}\frac{dn(x)}{dx}\frac{dx}{x^2}
\end{equation}
 The contribution from the constant cross-section above 30 MeV is found to be low. We have considered  a disc shaped region centered at the Galactic center of radius $R=10$kpc and height $z=0.5$ kpc having uniform intensity of radiation. 
The rate of photo-disintegration is highest for Fe and decreases with decreasing A as shown in our Fig.1. The spectrum of neutrons [GeV$^{-1}$cm$^{-3}$ sec$^{-1}$] produced in photo-disintegration is 
\begin{equation}
\frac{dN_{n_s}(E_{n_s})}{dE_{n_s}\,dt\,dV}=
0.5 R_{ph}(E_{n_s}) \frac{dN_{A}(E_{n_s})}{dE_{n_s} \, dV} 
\label{neutron_flux}
\end{equation}
 Both neutrons and protons are stripped out from the nuclei. We have assumed $50\%$ of the stripped nucleons are neutrons and denote them by $n_s$. The energy of the stripped neutrons is assumed to be the same as that of the parent nucleons.
 The steady state density of cosmic ray nuclei is $\frac{dN_{A}(E_n)}{dE_n \, dV}$ expressed in per nucleon energy in eqn.(\ref{neutron_flux}). 
This density flux (GeV$^{-1}$ cm$^{-3}$) is 
\begin{equation}
\frac{dN_{A}(E_n)}{dE_n\,dV}=10^{-4} 
\frac{4\pi}{c (cm/sec)}\frac{A \,dN_{A}(E_{A})}{dE_{A}\,dS\,dt\,d\Omega}
\label{cr_density}
\end{equation}

The neutron flux [GeV$^{-1}$ cm$^{-2}$ s$^{-1}$ sr$^{-1}$] is calculated for the disc shaped region. The geometrical correction is done following the formalism discussed in paper \cite{joshi}.
\begin{equation}
J_{n_s}(E_{n_s})=\frac{1}{\Omega_G}\int \frac{dV}{4\,\pi\, r^2} \frac{dN_{n_s}(E_{n_s})}{dE_{n_s}\,dt\,dV}
\label{neutron_fA}
\end{equation}
 $r$ is the distance of the site of neutron production from the earth.
 The geometry of our Galaxy is symmetrical around the Galactic center. 
We have calculated the effective radius of the halo with respect to the observer on earth assuming a homogeneous spherical halo centered at the earth. This effective distance or radius is obtained after integrating over r, $R_{eff}=\int dV/(4\,\pi\,r^2)=$1.7 kpc as discussed in \cite{joshi}. 
\par
The positions of the Galactic astrophysical objects are shown in the skymap in Fig.1. The solid lines show the region $|b|<10^{\circ}$ which includes most of 
the photon emitting objects. We have selected an even smaller region $|b|<5^{\circ}$ to define the Galactic plane region in the present work. In this region the intensity of starlight is expected to be very high. The solid angle subtended by this region to us is $\Omega_G=$1 sr.
\begin{figure*}[t]
\centerline{\includegraphics[width=4in,angle=-90]{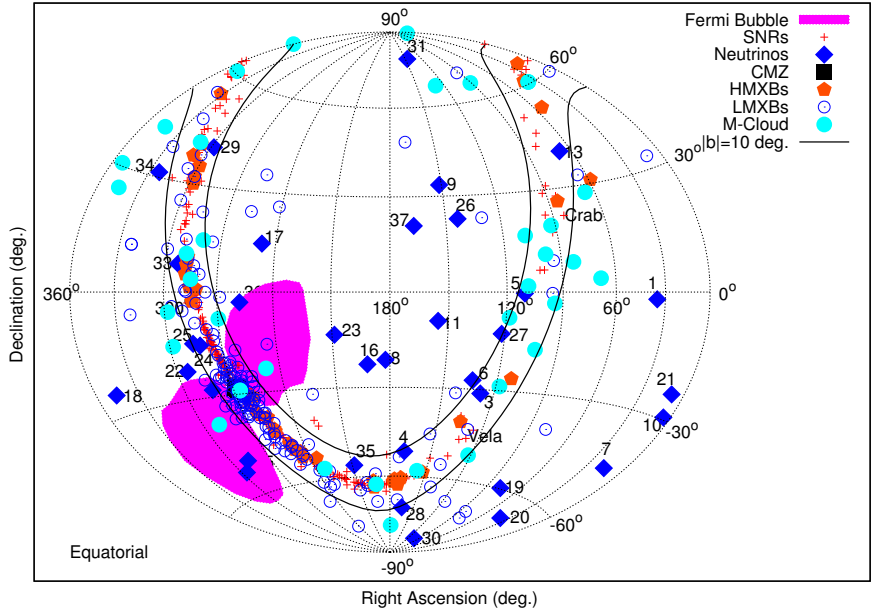}}
\caption{The positions of the Super Nova Remnants (SNR), Low and High Mass X-ray Binaries (LMXB, HMXB) from http://vizier.cfa.harvard.edu, Central Molecular Zone (CMZ), Galactic Molecular clouds (M-cloud) from https://www.cfa.harvard.edu are plotted in this skymap. The IceCube detected events (diamond shaped points) \cite{aart4} and the Fermi Bubbles (two patches) are also shown \cite{acker}.}
\label{fig:skymap}
\end{figure*}

The neutrons decay to antineutrinos with average energy $\epsilon'_n=0.48$ MeV in the rest frame of the neutrons. 
The secondary neutrons have a flux given by eqn.(\ref{neutron_fA}).
 The antineutrino flux (GeV$^{-1}$ cm$^{-2}$ sec$^{-1}$ sr$^{-1}$) produced from the decay of neutrons is 
\begin{equation}
J_{\bar\nu}(E_{\bar\nu})=\frac{m_n}{2\epsilon'_n}\int_{\frac{m_n E_{\bar\nu}}{2\epsilon'_n}}^{E_{n_s}^{max}}\frac{dE_{n_s}}{E_{n_s}} J_{n_s}(E_{n_s})
\label{anch_neuflux}
\end{equation}
as given in eqn.(25) of \cite{anch1} where $E_{n_s}^{max}$ is the maximum energy of the neutrons decaying to antineutrinos. The antineutrinos may take energy between $0$ and $\frac{2\epsilon'_n}{m_n} E_{n_s}$. We have assumed that the antineutrinos on average have energy $\frac{\epsilon'_n}{m_n} E_{n_s}$ to simplify the calculations. The number of neutrons is equal to the number of antineutrinos if all the neutrons decay before reaching us. We apply the conservation of the total number of neutrons and antineutrinos to get the antineutrino flux. The antineutrino flux (GeV$^{-1}$ cm$^{-2}$ sec$^{-1}$ sr$^{-1}$) produced from the decay
of neutrons of energy $E_{n_s}$ and Lorentz factor $\gamma_{n_s}$ is
\begin{equation}
J_{\bar \nu}(E_{\bar\nu})=\frac{m_n}{\epsilon'_n} J_{n_s}(E_{n_s})
\label{neu_flux}
\end{equation}
where $E_{\bar\nu}=\gamma_{n_s} \epsilon'_n$. We note that the antineutrino fluxes calculated using eqn.(\ref{anch_neuflux}) and eqn.(\ref{neu_flux}) differ by a factor of $2^{\alpha-1}/{\alpha}$ where $\alpha$ is the spectral index of the stripped neutron spectrum. For $\alpha$ in the range of 1 to 4 one gets accurate results within a factor of 2 using our eqn.(\ref{neu_flux}). 
 Our calculated antineutrino flux is found to be very low as shown in Fig.3.
\begin{table}
\begin{center}
\begin{tabular}{|c|c|c|c|c|c|c|}
\hline
& Parameters & p & He& CNO  &  Mg-Si & \begin{tabular}{c}
	                          $Fe$
 
\end{tabular}
 \\
\hline
Pop 1:& $a_{ij}$ & 7860& 3550& 2200 & 1430 & 2120  \\
$R_c=4$PV& $\gamma_{ij}$&1.66&1.58&1.63&1.67&1.63\\
\hline
Pop 2: &$a_{ij}$ &  20& 20 & 13.4 & 13.4 & 13.4\\
$R_c=30$PV &$\gamma_{ij}$&1.4&1.4&1.4&1.4&1.4\\
\hline
Pop 3 &$a_{ij}$& 1.7 & 1.7 & 1.14& 1.14 & 1.14 \\
$R_c=2$ EV &$\gamma_{ij}$&1.4&1.4&1.4&1.4&1.4\\
\hline
\end{tabular}
\end{center}
\vspace{0.2 cm}

{TABLE 1 : Hillas Model: normalisation constants $a_{ij}$ and spectral indices $\gamma_{ij}$ and cut-offs in rigidity $R_c$ (Gaisser, Stanev\& Tilav 2013).}
\label{hillas_model}

\end{table}

 \begin{table}
\footnotesize  
\begin{center}
\begin{tabular}{|c|c|c|c|c|c|c|}
\hline
& Parameters & p & He& C  &  O & \begin{tabular}{c}
	                          $Fe$
 
\end{tabular}
 \\
\hline
Pop 1:& $a_{ij}$ & 7000& 3200& 100 & 130 & 60  \\
$R_c=120$TV& $\gamma_{ij}$&1.66&1.58&1.4&1.4&1.3\\
\hline
Pop 2: &$a_{ij}$ &  150& 65 & 6 & 7 & 2.3\\
$R_c=4$PV &$\gamma_{ij}$&1.4&1.3&1.3&1.3&1.2\\
\hline
Pop 3 &$a_{ij}$& 14 &  & &  & 0.025 \\
$R_c=1.3$ EV &$\gamma_{ij}$&1.4&&&&1.2\\
\hline
\end{tabular}
\end{center}
\vspace{0.2 cm}

{TABLE 2 : Global fit Model: normalisation constants $a_{ij}$ and spectral indices $\gamma_{ij}$ and cut-offs in rigidity $R_c$ (Gaisser, Stanev \& Tilav 2013).}
\label{Global_model}
\end{table}

\section{Photo-Disintegration of UHECRs in the Extragalactic Medium} 

 The formalism discussed in the earlier section is extended to calculate the antineutrino flux produced in the photo-disintegration of ultrahigh energy cosmic ray nuclei during their propagation from extragalactic sources to the earth.

\par
 It has been suggested recently that the photo-disintegration of some very high injected flux of extragalactic cosmic ray nuclei of energy between $10^{8.5}$ GeV/nucleon to $10^{9.5}$ GeV/nucleon by the far infrared background at 10 meV may explain the neutrino/antineutrino flux detected by the IceCube detector in the energy range of $10^{5.5}$ to $10^{6.5}$ GeV \cite{anch2}. The flux of the injected or parent ultrahigh energy cosmic ray nuclei is unknown in this paper. 
When the ultrahigh energy cosmic ray nuclei are photodisintegrated high energy protons and neutrons are produced. The high energy neutrons decay to high energy protons and antineutrinos. The author has assumed that only 1 to 10$\%$ of the high energy protons reach us due to shielding by the extragalactic magnetic field and their flux is the same as the observed proton flux in the energy range of $10^{8.5}$ to $10^{9.5}$ GeV. The antineutrino flux from the decay of the high energy neutrons is calculated in this paper by assuming that this neutron flux is 10 to 100 times higher than the observed proton flux in the energy range of $10^{8.5}$ to $10^{9.5}$ GeV. Due to this assumption an antineutrino flux of $\sim 10^{-8}$ GeV cm$^{-2}$ sec$^{-1}$ sr$^{-1}$ is estimated in \cite{anch2}.
  
\par
The approach followed in our paper is completely different. We have used the steady state or observed flux of ultrahigh energy cosmic ray nuclei with energy $10^{8.5}$ GeV/nucleon to $10^{9.5}$ GeV/nucleon to calculate the flux of the high energy neutrons produced in photo-disintegration. The antineutrino flux produced in the decay of these neutrons is subsequently calculated. 
Thus the normalisation used in our paper does not depend on any assumption or magnetic shielding.
\par
 We have assumed a spherical cosmological volume within which the cosmic ray sources are distributed isotropically. The isotropic UHECR flux within this volume is assumed to be similar to that predicted in the Hillas model. The heavy nuclei are photo-disintegrated in the ambient radiation field and antineutrinos are produced through the decay of the UHECR neutrons. At 10 meV the intensity of the radiation field is 1 eV/cm$^{3}$ near the Galactic center according to the radiation fields (IR1 and IR2) calculated in \cite{mos}. The intensity of the extragalactic background light at this energy has un upper limit of 0.005 eV/cm$^{3}$ from observations on gamma ray sources \cite{dwek}, which is 200 times lower than that at the Galactic center. 
The energy of the ambient low energy photons is boosted in the nuclear rest frame by the Lorentz factor of the nucleus which disintegrate the cosmic ray nucleus. The contribution from the GDR cross-section is maximal for incident photon energy in the energy range of 10-30 MeV in the nuclear rest frame.
The CMB is present throughout the interstellar medium. However, for photo-disintegration by CMB the UHECRs have to be of energy more than $10^{11}$ GeV/nucleon and the secondary antineutrinos would have energy more than $10^{8}$ GeV.  At present we do not have any observational result on the neutrino/antineutrino flux at this energy. 
\par
Our Eqn.(10) has been modified to find the flux in the observer's frame 
\begin{equation}
J^{ob}_{n_s}(E^{ob}_{n_s})=\frac{1}{4\pi} \int \frac{dV_c}{4\pi d_c^2} \frac{dN_{n_s}(E_{n_s})}{dE_{n_s}dt dV_c}.
\label{neutronf_cosm}
\end{equation} 
The comoving distance and differential volume are $d_c$ and $dV_c$ respectively. The effective radius of the cosmological volume centered at the observer on earth is $R_{eff}=\int dV_c/(4\pi d_c^2)$. Within this volume the flux of the ultrahigh energy cosmic ray nuclei is assumed to be the same as their observed flux near the earth. The densities of the radiation field and the ultrahigh energy cosmic rays are assumed to be uniform within this volume. If the redshift corresponding to the distance $R_{eff}$ is $z_{eff}$ then the energy and time in the observer's frame and at redshift $z_{eff}$ are related as
$E^{ob}=E/(1+z_{eff})$, $t^{ob}=t(1+z_{eff})$. The antineutrino flux per unit energy unit time in the observer's frame is
\begin{equation}
J^{ob}_{\bar\nu}(E^{ob}_{\bar\nu})=\frac{m_n}{\epsilon'_n} J^{ob}_{n_s}(E^{ob}_{n_s})
\label{frame_corr}
\end{equation}
 
 Eqn.(\ref{neutronf_cosm}) and Eqn.(\ref{frame_corr}) are used to estimate $R_{eff}$. The AGN which are bright in X-rays and potential sources of extragalactic cosmic rays, are located near redshift 2 to 3 \cite{miyaji}. Assuming $z_{eff}\sim 2$ the correction to the energy of the antineutrinos in the observer's frame can be neglected. To produce the observed flux of $6.7\times 10^{-18}$ GeV$^{-1}$cm$^{-2}$ sec$^{-1}$ sr$^{-1}$ near $10^{5}$ GeV \cite{ice_new} the required effective radius is found to be unrealistically high $\sim 50000$ Gpc. As the density of the UHECR nuclei is very low  a large effective radius or volume is required to explain the flux detected by the IceCube detector.

\section{Discussion:}
We have used the three population models (i) Hillas model (ii) global fit model 
 of cosmic rays to calculate the flux of antineutrinos produced in the photo-disintegration of cosmic ray heavy nuclei near the Galactic plane region.
The CNO, Mg-Si and Fe cosmic ray nuclei are included in our study.
The radiation field is assumed to be uniform within the disc shaped region of radius 10 kpc and height 0.5 kpc centered at the Galactic center. The spectral intensities of the field corresponding to R=0,z=0 and R=4kpc, z=0 given in Fig .1. (bottom panel) of the paper by \cite{mos} have been used in the present work. 
\par
 The cross-section of photo-disintegration has a resonance below 30 MeV and above 30 MeV it is a constant with a smaller value. We have calculated the rates of photo-disintegration of oxygen nuclei in the radiation field IR1 for the resonance  and constant values of the cross-section. The contribution from the constant cross-section is found to be low.
Our calculated rates of photo-disintegration for oxygen and iron nuclei ($R_{ph}$) are plotted in Fig.2. for the radiation fields defined at R=0,z=0 (IR1) and at R=4kpc,z=0 (IR2). 
The energy dependence of the rate of photo-disintegration 
can be explained from the energy dependence of the background radiation field.
The higher energy cosmic ray nuclei can be photo-disintegrated by the lower energy photons in the radiation field.
\par 
If we increase the halo size the effective radius increases (please see Table 1: of \cite{joshi}). The antineutrino flux is directly proportional to the effective radius so it also increases. 
 \par
 The IceCube collaboration has measured the isotropic flux of neutrinos and antineutrinos in the TeV-PeV energy range. In the energy range of 25 TeV to 2.8 PeV their measured flux is $E_{\nu}^2 \frac{dN(E_{\nu})}{dE_{\nu}}=6.7\times 10^{-8} (E_{\nu}/100 TeV)^{-0.5}$ GeV cm$^{-2}$ sec$^{-1}$ sr$^{-1}$ \cite{ice_new}. 
 Our calculated fluxes from the Galactic plane region shown in Fig.3. are many orders of magnitude lower than the flux detected by the IceCube detector. The average length traversed by the neutrons before decaying to antineutrinos increases with energy. At $10^8$ GeV this length is 1 kpc. We do not expect antineutrinos from the Galactic plane region at very high energy as their parent ultrahigh energy neutrons will not decay before reaching us.
\par

 After showing that the antineutrino flux produced in the Galactic plane region is very low we have extended our formalism to the extragalactic medium. 
Assuming the sources are located at a effective redshift 3 we have calculated the antineutrino flux from photo-disintegration of the flux of UHECR nuclei in the global fit model by the extragalactic background light. The upper limit on the intensity of extragalactic background light has been used from \cite{dwek} in our calculation. Our calculated antineutrino flux is shown in Fig.4.
In this case also the contribution to the observed flux from the photo-disintegration of UHECR nuclei is very low.
\begin{figure*}[t]
\centerline{\includegraphics[width=2.2in,angle=-90]{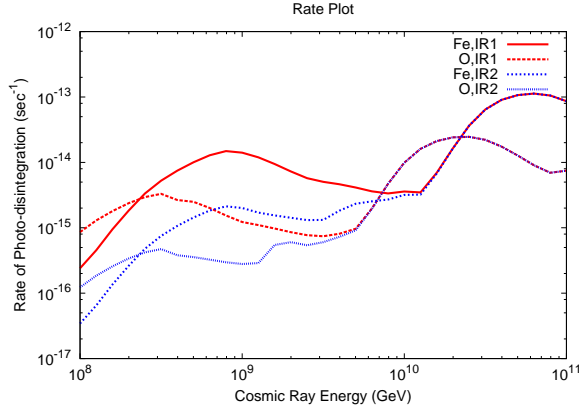}}
\caption{Rate of Photo-disintegration of cosmic ray iron and oxygen nuclei 
plotted for the assumed radiation fields IR1 (R=0,z=0) and IR2 (R=4kpc,z=0).}
\label{fig:rateplot}
\end{figure*}
\begin{figure*}[t]
\centerline{\includegraphics[width=2.2in,angle=-90]{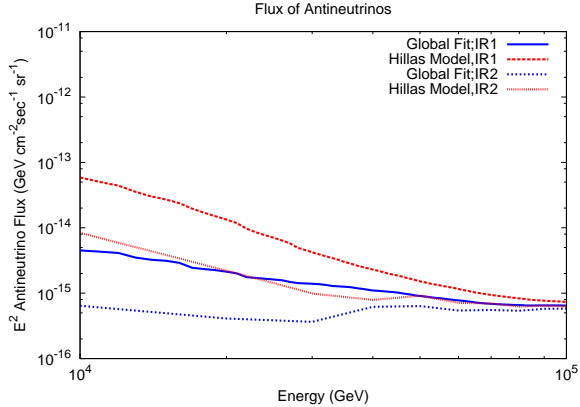}}
\caption{Antineutrino fluxes from the Galactic plane region calculated in the energy range of 10 TeV to 100 TeV for the Hillas model and the global fit model assuming uniform radiation fields for the two cases IR1 and IR2.}
\label{fig:antineu}
\end{figure*}
\begin{figure*}[t]
\centerline{\includegraphics[width=2.2in,angle=-90]{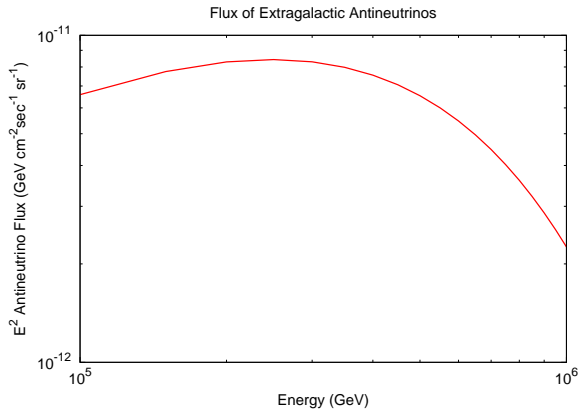}}
\caption{Maximum flux of antineutrinos calculated for effective redshift of sources $z_{eff}=3$ and maximum extragalactic background light 0.005 eV/$cm^3$ \cite{dwek}. The global fit model of the UHECR flux has been used. }
\label{fig:antineu}
\end{figure*}

\section{Conclusion:}
The three population models are used to calculate the antineutrino flux produced in the photo-disintegration of CNO, Mg-Si and Fe nuclei near the Galactic plane region. In all cases it is found that the antineutrino flux is much less than the flux measured by the IceCube detector. The antineutrino flux is directly proportional to the effective radius of the spherical region within which the cosmic ray nuclei are photo-disintegrated. The formalism discussed for the Galactic plane region has been extended to study the contribution from the photo-disintegration of cosmic ray nuclei outside our Galaxy. In this case also the antineutrino flux is very low compared to the observed flux due to the low flux of the parent UHECR nuclei.
\section{Acknowledgment}
The author is thankful to Jagdish Joshi and Kumar Ravi for providing the skymap.

\end{document}